\documentclass[review]{elsarticle}

\usepackage{lineno,hyperref}
\modulolinenumbers[5]

\usepackage{graphicx}

\usepackage[cmex10]{amsmath}
\usepackage{amssymb}

\usepackage{algorithmic}

\newcommand{\hl}[1]{{#1}} 

\begin{document}

\begin{frontmatter}

\title{The equal load-sharing model of cascade failures in power grids}

\author[cnr,imt,lims]{Antonio Scala\corref{mycorrespondingauthor}}
\cortext[mycorrespondingauthor]{Corresponding author}
\ead{antonio.scala@cnr.it}
\ead[url]{https://sites.google.com/site/antonioscalaphys/}

\author[gubkin]{Pier Giorgio De Sanctis Lucentini}

\address[cnr]{ISC-CNR Physics Dept., Univ. dir Roma "La Sapienza", 00185 Roma, Italy}
\address[imt]{IMT Alti Studi Lucca, piazza S. Ponziano 6, 55100 Lucca, Italy}
\address[lims]{LIMS London Institute of Mathematical Sciences, 22 South Audley St Mayfair London UK}
\address[gubkin]{Gubkin Russian State University of Oil and Gas, Moskow, Russia}

\begin{abstract}
Electric power-systems are one of the most important critical infrastructures. In recent years, they have been exposed to extreme stress due to the increasing demand, the introduction of distributed renewable energy sources, and the development of extensive interconnections. We investigate the phenomenon of abrupt breakdown of an electric power-system under two scenarios: load growth (mimicking the ever-increasing customer demand) and power fluctuations (mimicking the effects of renewable sources). 
Our results indicate that increasing the system size causes breakdowns to become more abrupt; in fact, mapping the system to a solvable statistical-physics model indicates the occurrence of a first order transition in the large size limit. Such an enhancement for the systemic risk failures (black-outs) with increasing network size is an effect that should be considered in the current projects aiming to integrate national power-grids into "super-grids".
\end{abstract}


\end{frontmatter}


\section{Introduction}

The electrical power system (EPS) is crucial for the well functioning of  most critical infrastructures like telecommunications, banking systems, oil  and gas pumping, water distribution \cite{Rinaldi2001}. 
Since their first appearance in 1881 at Godalming in England, EPSs  have evolved into one of the most well engineered and robust network infrastructure; nevertheless power outages do occur with a likelihood  larger than what would be naively expected. In particular, historical data show that outages' empirical probability distribution  has fat tails \cite{EPRIrep2005,Rosas-Casals2011}, corresponding to a non-vanishing risk of system-wide failures (major outages or black-outs) causing disruption and economic damages. 
  
Black-outs often occur as a cascading sequence of failures and automatic disconnections triggered by an apparently minor initiating event; no two cascading outages are the same \cite{CanUSblackoutRep2003}.
The mechanisms driving the tripping (disconnections) sequences are manifold and comprise the overloading of line and/or generators, frequency imbalances and transient currents. A black-out happens when the time-scales of automatic reactions are way to fast to correct the process by human intervention. 

In order to predict and control the system, power engineers have developed and are refining sophisticated systems to simulate the full dynamics of whole power systems; moreover, distributed metering is going to furnish detailed data for the energy consumption at the household level.  
Nevertheless, black-outs still occur and understanding the nature of such occurrences is still an open problem. A general question is whether such large outages can be partially due to emergent behaviour in the EPSs: if this would be the case, increasing the accuracy of power systems' simulation would not result in better predictions of black-outs. The fact that EPSs are aggregations of large number of simple units makes them an ideal candidate to be a system exhibiting additional complexity as a whole beyond what is dictated by the simple sum of its parts.

To highlight the possibility of emergent behaviour, it is first necessary to simplify the system in order to understand the basic mechanisms that could drive systemic behaviour. We will hence introduce a simplified model of EPS that is amenable of both simulation on realistic grids and of a self-consistent analytical solution. We will then consider the behaviour of such a model under two kinds of stress: (1) an increasing growth of the loads, mimicking the case of EPS that are operated to the limit of their capacities in order to maximize profits and (2) fluctuations in demands and generation, mimicking the effects of the steady penetrations of the erratic renewable sources. This latter case if of particular interest since the effects and consequences of introducing in the grids new erratic sources have not yet been fully understood.

\section{Overload Cascade Model}

A particular source of stress to EPS comes to the fact that adjustments in 
power generation are not real-time but follow fixed time schedules; for 
example, in Europe the production is fixed in advance the day before and 
periodic adjustments happen every $15$ minutes; therefore, the reaction time 
(apart from automatic controls/tripping) is generally much higher than the time 
of propagation of electrical perturbation in the system.

The tripping of lines and generators above their operating limits induced by 
automatic protective equipments is common to all kinds of cascading outages; 
while this process is intended to protect costly equipments from damage, it can 
potentially widen cascade failures \cite{CanUSblackoutRep2003}. 

Most of the cascade model for power grids are purely topological models based  on the local redistribution of power loads upon failure \cite{CrucittiPRE2004,SolePRE2008,WangSS2009,WangSS2011} and disregard the long range  nature of electricity. 
On one hand, a clear signature of the non-locality of power outages can be found in real data: this is, for example, the case of the tripping sequence of the Western Interconnection WSCC system disturbances in July 2-3, 1996 \cite{DisturbancesNERC1996}, where the occurrence of subsequent failures in far away lines can be observed. 
On the other hand, it is possible to have simple yet realistic models respecting Kirchoff laws, 
like the overload cascade model (OCM) introduced by Pahwa et al \cite{Pahwa2010}. 

In Pahwa's model, an initial power flow configuration is calculated using the DC power flow model\cite{GungorBOOK1988}; link capacities (i.e. the maximum of power-flow before a link failure) is proportional to the initial flow on the branch and is usually assumed to be $10\%$ higher than the initial flow. Notice that this condition is equivalent to the normal bounds assumed for the validity of the DC power flow model. 
To understand what happens when the system is subject to stress, a new configuration of loads is assumed and power flows recalculated. If the load on a line goes beyond its capacity, the line trips (disconnects) and power flows are recalculated on the new topology (i.e. the grid MINUS the tripped lines). Such procedure is repeated until convergence. Notice that OCM is a simplified version of the dynamic OPA model introduced by Carreras et al \cite{CarrerasChaos2002} that is also based on DC power flow equations but uses linear optimization for generation dispatch after each link failure step. \hl{Further simulation evidence of the nature of a phase transition associated to load increases can be found in} \cite{Chen2005,Nedic2006}.

\section{Equal load-sharing model}

The OCM model, as several similar ones, is based on DC power flow's equations. In the DC power flow, the grid is described by matrix $Y_{ij}$ measuring the the admittance of the $M$ edges (branches) $ij$ among $N$ nodes (buses). The power injected (generators) or absorbed (loads) on the $i$-th bus is described by the vector $P_i$.  The variables describing the electric state of the systems are the phase angles $\theta_i$ at the buses that are linearly related to the injected powers: 
\begin{equation}
\mathcal{L}\vec{\theta}=\vec{P}
\label{eq:DC}
\end{equation}
where $\mathcal{L}$ is the Laplacian associated with the admittance
matrix $Y$. Power flows along branches are proportional to the phase-angle differences $\theta_{ij}=\theta_{i}-\theta_{j}$.

Since eq. (\ref{eq:DC}) is in the form of a Laplacian, the interactions among phase angles $\theta_i$ is long range; in fact, similar equations hold to describe resistor network models \cite{deArcangelis1985} or simplified fracture models \cite{ZapperiPRL1997}. Hence, when an edge breaks down in such kind of systems, the stress it was carrying (in our case the load flow) will redistribute among all the remaining edges. To capture the above effect, we can hence develop a mean-field model of the OCM by assuming that the load carried by a tripped line gets equally shared among all the remaining links: we call such a model the Equal Load-Share model (ELSM) of cascading failures in power grids. Notice that such an approximation corresponds to the one of the so-called democratic fiber-bundle model of fractures \cite{Peirce1926,Daniels1945}. 

In the ELSM, we consider $M$ lines where each line $e$ is characterised by its capacity (i.e. the maximum amount of power flow) $C_{e}$; by simplicity, we assume that such capacities are characterised by a probability distribution $p\left(C\right)$. Also the initial stresses $l_e$ on the lines can be characterised by a probability distribution  $q\left(l\right)$.
The ELSM dynamics then follows the same dynamics of the OCM: when the load of a line surpasses its capacity, the line breaks; the characteristic of the ELSM is that such load gets equally redistributed among all the remaining surviving links.

\subsection{Uniform Stress}

We first consider a uniform stress situation, where the system is subject to an overall load $L$ and all the links are initially subject to the same load $l^0=L/M$. The fraction of links $f^{1}=\int_{0}^{L/M}p\left(C\right)dC$ immediately fails since their thresholds are less than the load $l$ they sustain. Hence, after the first stage of a cascade, there are $M^{1}=(1-f^{1})M$ surviving links and the new load per link is $l^{1}=L/M^{1}$. The following cascade's stages follow analogously; we can thus write the mean field equations for the $\left(t+1\right)$-the stage of the cascade:
\begin{equation}
\left\{ \begin{array}{lll}
l^{t+1}&=&L/M^{t}\\
f^{t+1}&=&P\left(l^{t+1}\right)\\
M^{t+1}&=&\left(1-f^{t+1}\right)M
\end{array}\right.
\label{eq:MFcascadeStages}
\end{equation}
with initial conditions $l^{t=0}=L/M$, $f^{t=0}=1$, $M^{t=0}=M$ and
where $P\left(x\right)=\int_{0}^{x}p\left(C\right)dC$ is the cumulative
distribution function associated to $p(C)$. Eq.$\,$\ref{eq:MFcascadeStages} 
can be simplified in a single equation for the fraction $f$ of failed links
\begin{equation}
f^{t+1}=P\left(\frac{L}{\left(1-f^{t}\right)M}\right)
\label{eq:FailureFractionIteration}
\end{equation}

and the fix-point
\begin{equation}
f^*=P\left[\frac{L}{\left(1-f^*\right)M}\right]
\label{eq:FailureFractionFixpoint}
\end{equation}

represents the total fraction of links broken by the failure cascade. The behaviour of $f^*$ depends on the functional form of $p\left(C\right)$ \cite{daSilveiraPRL1997} and is known to present a first order transition for a wide family of curves. \hl{Notice that the total
load stays constant during the cascades process, while in real networks total load decreases as soon as link failures disconnect a node asking power. Moreover,due to the mean-field nature of the approximation, no islanding phenomena are considered.}

To mimic the functional form of the distribution for the thresholds $C$ in the case of realistic power networks, we analyse the OCM thresholds' frequency distribution for several IEEE networks. 
We find that such histograms can be approximated by $p$'s of the form
\begin{equation}
p\left(c\right)\sim c^{-\gamma}\,\,\mbox{ for }\,\,c_{min}<c<1
\label{eq:p_of_c}
\end{equation}
where $c$ are the normalised line capacities $c=C/C_{max}$ and $C_{max}$ is the maximum line capacity of the network. For all the networks we find that $c_{min}\sim0.02$ and that the exponent $\gamma$ increases according to the size.  
In fact, in \cite{PahwaSciRep2014} it is show that the histograms for the IEEE 14 bus network have approximately a flat distribution with $\gamma \sim 0$, while already for the IEEE 247 bus network the distribution has a sharp decrease with $\gamma \sim 2$. Notice that the distribution described by eq. \ref{eq:p_of_c} are not truly power-law distributions since the maximum line capacity $C_{max}$ is not growing with the system size since is limited by the maximum power a real cable can hold. \hl{Notice that since power grids are topologically sparse graphs, i.e. the number of edges $M$ is proportional through a small factor (the average degree) to the number of nodes $N$ }\cite{AmaralPNAS2000}\hl{, both such quantities can be used as a proxy for the system size.}
 
\begin{figure}[!H]
\centering
\includegraphics[width=3.5in]{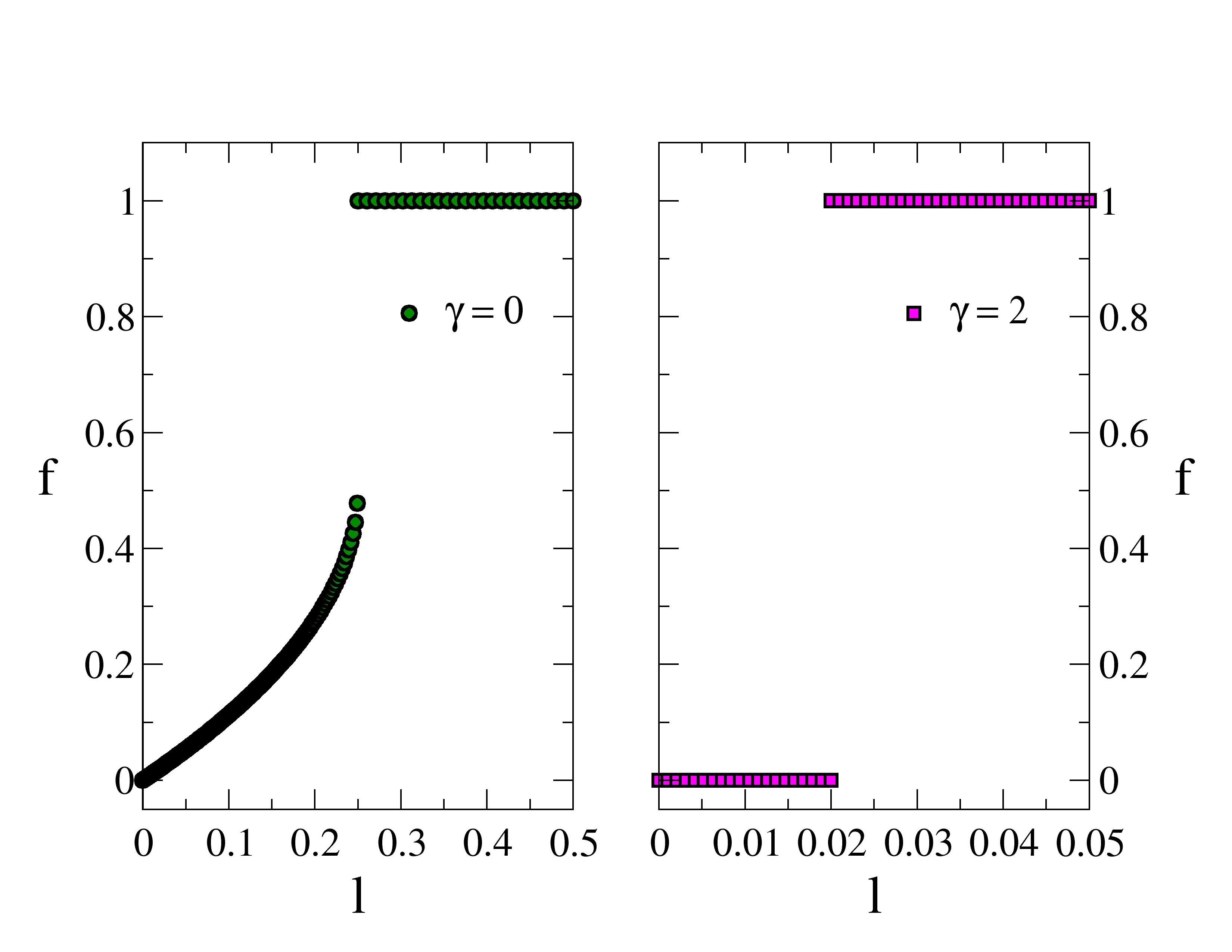}
\caption{ELSM first order transition in the fraction of failed links calculated via eq. (\protect{\ref{eq:FailureFractionFixpoint}}) versus the initial overall load of the system.
Left panel: the case of uniform line capacities $p\left(c\right) = const$.
Right panel: the case of line capacities distributed according to $p\left(c\right) = c^{-2}$.  \label{fig:FlatVsPLaw}}
\end{figure}

In fig.$\,$\ref{fig:FlatVsPLaw} we show the results of the ELSM by plotting the fix-point of equation \ref{eq:FailureFractionFixpoint}) for the extreme cases of a flat distribution ($\gamma=0$) and of a sharp decrease in the distribution ($\gamma=2$). 
In both cases we find a marked jump due to the first order of the transition. We observe that higher values of $\gamma$ makes the network more fragile, i.e. there is a bigger jump and less links break before the transition; moreover, the system breaks down at a lower stress threshold $l^*$. Such a behaviour is to be expected since for larger $\gamma$ there is an higher fraction of links with a very low threshold that are going to fail all together at the initial stages of the cascades and will inflict a higher damage than the case of a uniform distribution $\gamma = 0$. Such effect could indicate that interconnecting power networks could increase their fragility; in fact, such process often corresponds to adding few high capacity lines, i.e. "lowering the tail" of the normalised capacity distribution $p\left(c\right)$.

An artificial effect due to the mean-field spirit of the model is the fact that, after the threshold $l^*$, the system breaks down completely ($f=1$). This is not to be expected in realistic grids, where the system can often break down in islands \cite{PahwaSciRep2014}.

Notice that the ELSM shares the same mean-field spirit of the CASCADE model for black-outs \cite{DobsonHAWAI2002,DobsonPEIS2005}; while the ELS model assumes a starting situation in which thresholds are randomly distributed, CASCADE assumes that the thresholds are the same. As shown by \cite{daSilveiraPRL1997}, the different behaviour of ELSM and CASCADE (presenting respectively a first order and a second order transition) can be imputed to the different functional form of the $p\left(C\right)$ for the two models. However, recent analysis on black-outs' sizes in several countries indicates that causal factors other than the self-organization or a critical state predicted by CASCADE might be significantly ruling the system dynamics \cite{Rosas-Casals2011}.
\subsection*{Random stress}

The case in which there are also fluctuations in the initial loads cannot be described by a single equation for average quantities: in fact, the failure of a line $e$ will depend on both its capacity and the specific stress applied. However, the fraction $f$ of broken links for 
such a case can be solved numerically by iterating the following algorithm:

\begin{flushleft}
\begin{algorithmic}
\STATE $\mathcal{T} \leftarrow \emptyset$ (set of tripped lines)
\STATE $S \leftarrow 0$ (magnitude of redistributed stress)
\REPEAT
\STATE $\Delta S \leftarrow 0$
\STATE $M_a \leftarrow M- \vert \mathcal{T} \vert$ (number of active lines)
\FORALL{$l_e+S/M_a>C_e$ , $e \in \complement \mathcal{T}$}
\STATE $\Delta S \leftarrow \Delta S + l_e$
\STATE $\mathcal{T} \leftarrow \mathcal{T} \cup \left\lbrace e \right\rbrace$
\ENDFOR
\STATE $S \leftarrow S+ \Delta S$ 
\UNTIL{$\Delta S = 0$}
\label{alg:rndELS}
\end{algorithmic}
\end{flushleft}

\begin{figure}[!H]
\centering
\includegraphics[width=3.5in]{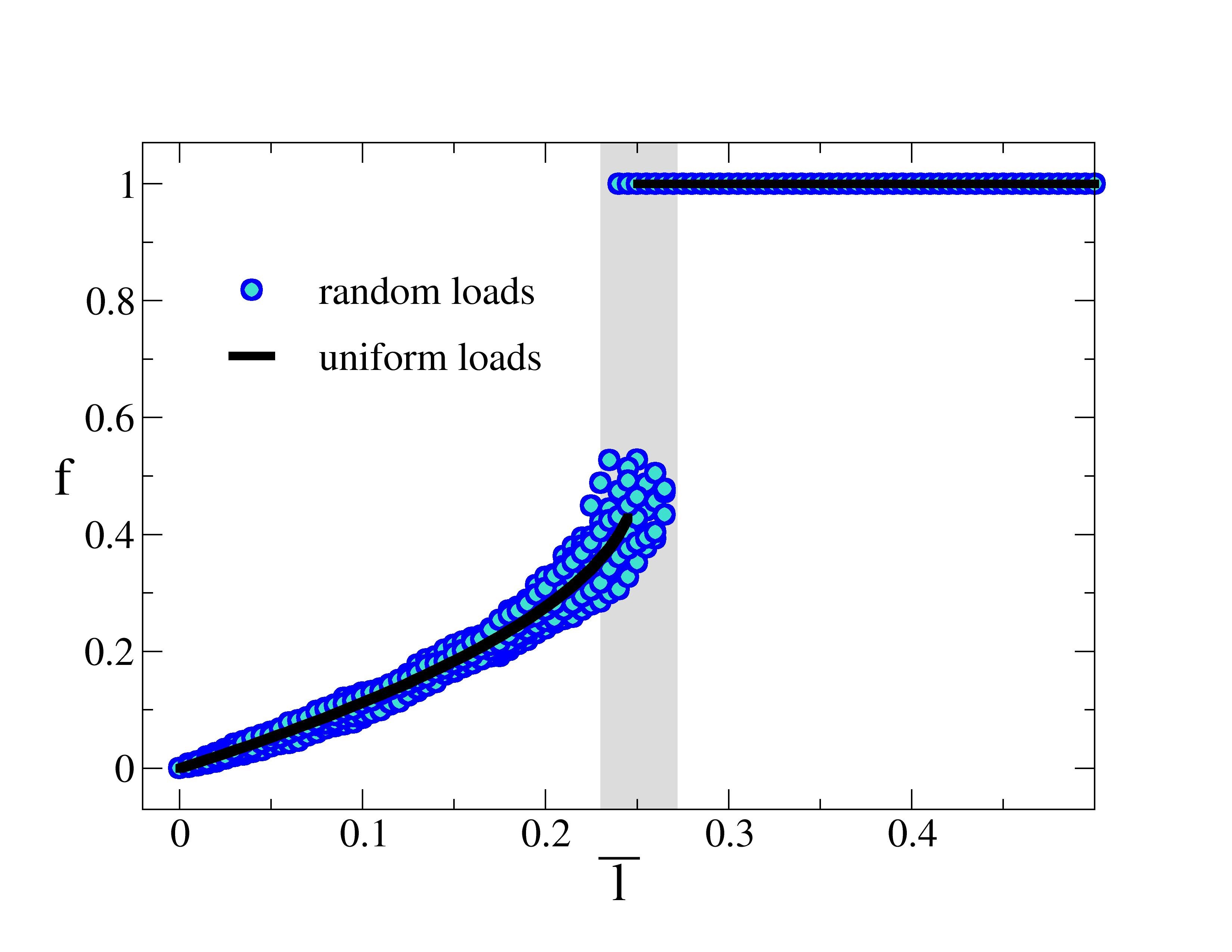}
\caption{  
Behaviour of the fraction $f$ of failed links versus the average stress $\overline{l}$ 
for $10$ realizations of the disorder in a system with $M=10^3$ links
(simulations follow the algorithm of sec.\textsl{III}.
Notice that the breakdown of the system does not happen at a single point but 
in a region around the critical point of the non-disordered system.
\label{fig:rndLoads}}
\end{figure}

For the sake of simplicity, we will consider a case in which the initial loads $l_{e}^{0}$ are uniformly distributed among $0$ and $2\overline{l}$, i.e. the fluctuations $\sigma = \overline{l} / \sqrt{3}$ of the initial loads are of the same order of magnitude the average initial load $\overline{l}$. Notice that the case $\sigma=0$, $\overline{l}=L^0/M$ recovers the previous case solved by eq. (\ref{eq:FailureFractionIteration}).  
By solving numerically such system at finite $M$ for different several realizations of the initial loads, we find that the transition does not happen any more
at a critical value of $\overline{l}$ but can occur in a region
around the mean-field prediction. As an example, we show in fig.$\,$\ref{fig:rndLoads}
the results for $10$ realizations of the initial loads in a
system of $M=1000$ links. 

\begin{figure}[!H]
\centering
\includegraphics[width=3.5in]{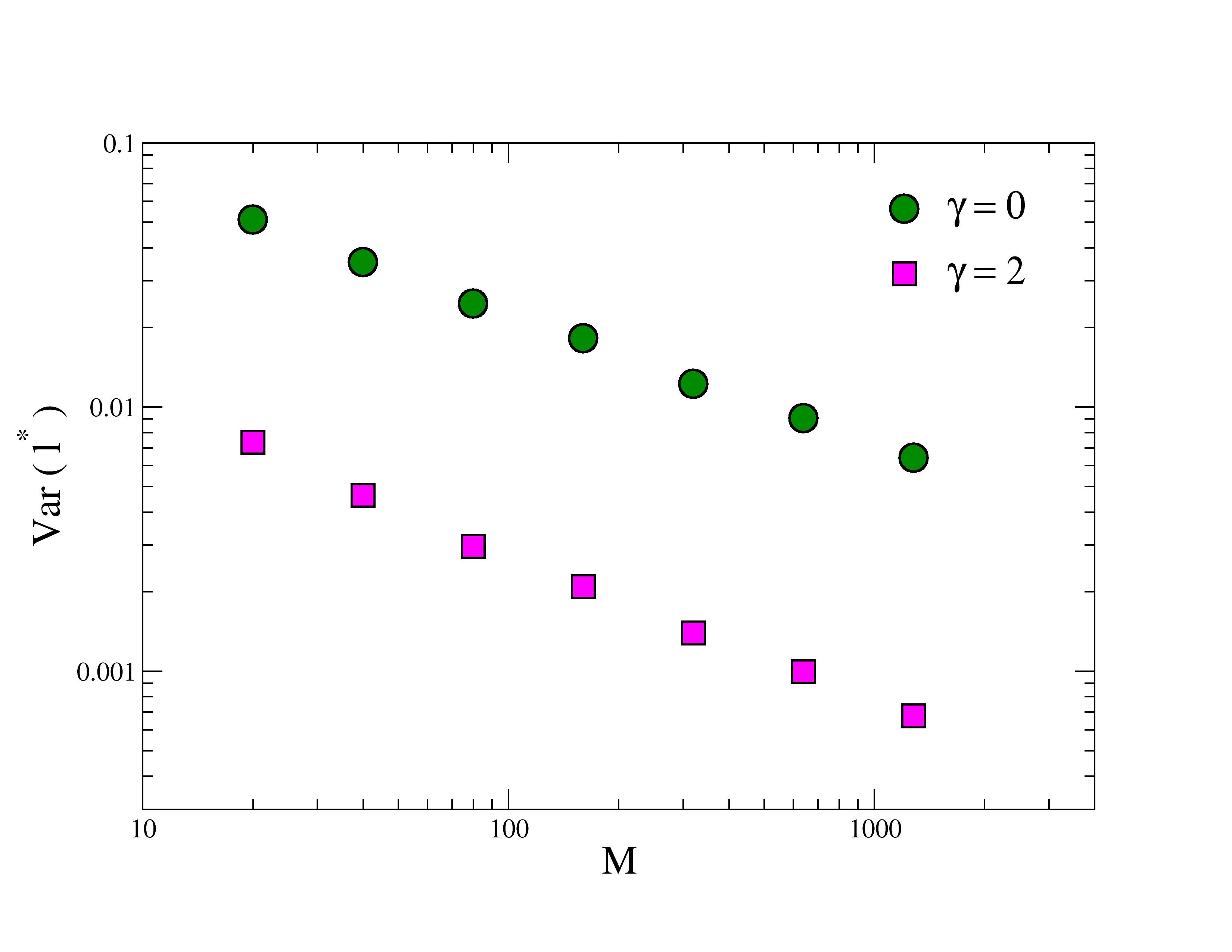}
\caption{The width $W$ of the region where the random system can break down 
can be measured by the variance 
$\sqrt{\left\langle\left(l^*\right)^2\right\rangle-\left\langle l^*\right\rangle^2}$
of the set of point where the system breaks down when submitted to random stresses.
We find that the width $W$ goes to zero approximately as $M^{-1/2}$; hence in bigger 
systems the transition region shrinks toward a single point.
\label{fig:VarGoesTo0}}
\end{figure}

We notice that in presence of disorder makes the transition point at which the abrupt cascade happens can vary inside a region. To investigate such effect, we investigate the variation of the size of such a region with the system size, \hl{where we use the number of edges $M$ as a measure of the system size}. We find that the distribution of failure points (i.e. the critical loads at which the system breaks down) follows a Gaussian distribution in presence of disorder. Hence, we identify the width $W$ of the transition region with the variance $Var\left(l^*\right)=\sqrt{
\left\langle\left(l^*\right)^2\right\rangle-\left\langle l^*\right\rangle^2
}$ of the transition point. In fig.$\,$\ref{fig:VarGoesTo0} we show that the 
width $W$ calculated over several configuration of disorders goes to zero with 
the system size $M$ approximatively as $M^{-1/2}$.

\section{Conclusion}
Statistical mechanics represent a useful framework for studying and analysing structure, dynamics and evolution of many complex systems\cite{RosasCasalsIJCIS2015}.
In this paper we have introduced a solvable model of cascading failure of power grids that captures the long-range interactions of the electric system. Such model predicts that cascading failures are a first order phenomena, i.e. the system abruptly breaks down when stressed.
Such abrupt failures are very difficult to predict since they often come without showing precursor signs.
Moreover, we analise such model in presence of randomness in the stresses: in such a case, we find that the abrupt breakdown does not happen at a single point but can happen in a whole region where the system is unpredictable, i.e. can be either safe or totally wrecked depending on the actual stresses applied to the system. Nevertheless, the width of such region seems to go to zero with the system size \hl{$M$} in our mean-field approximation.
As a limitation, the mean-field character of the model fails to capture the possible fragmentation (akin to the programmed islanding used to isolate large black-outs) that can stop cascading failures in real systems and planar networks. Given the difficulty of predicting such ruptures in the system, a possible approach would be to introduce dynamic strategies in the system like the self-healing procedures outlined by Quattrociocchi et al. \cite{QuattrociocchiPONE2014}.

\section*{Acknowledgement}

AS acknowledges the support from 
CNR-PNR National Project ”Crisis-Lab”,
EU FET project DOLFINS nr 640772,
EU FET project MULTIPLEX nr.317532
and EU HOME/2013/CIPS/AG/4000005013 project CI2C.
The contents of the paper do not necessarily reflect the position or the policy 
of funding parties. 

During the review of the paper, we have discovered that Yagan has solved a similar model in the case of a finite fraction $p$ of line failures$\,$\cite{YaganPRE2015}.

\end{document}